\documentclass[twocolumn,10pt]{revtex4-2}

\begin{document}

\newcommand{\be}{\begin{equation}}
\newcommand{\ee}{\end{equation}}
\newcommand{\bea}{\begin{eqnarray}}
\newcommand{\eea}{\end{eqnarray}}

\title{(Pre-) Modern (Non-) Fermi Liquids}

\author{D. V. Khveshchenko} 
\affiliation{Department of Physics and Astronomy, 
University of North Carolina, Chapel Hill, NC 27599}

\begin{abstract}
\noindent
This note addresses the problem of constructing a proper bosonized description of the collective modes in strongly interacting (non-)Fermi liquids which is specific to two spatial dimensions.
Although, in a mild form, this subtlety exists in the Fermi liquid as well, the discussion focuses on
the effects of long-ranged and/or retarded interactions which can completely destroy the fermionic quasiparticles. The present analysis also provides a further insight into the nature and properties of the collective bosonic modes in such systems. 

\end{abstract}

\maketitle


The unusual properties of the so-called non-Fermi liquids (NFL) have been studied
by a rather diverse variety of methods and techniques over several decades. 

Some authors elaborated on the earlier renormalization-group analyses \cite{rg} in the hope of extending them 
to the poorly (or even totally un-) controllable regime of non-weak interactions by 
introducing some (arguably, rather unphysical) small parameter, such as a deviation from some (fractional) critical dimension \cite{sslee}. 
Apart from reproducing the common layout of the conjectured phase diagram, though, thus far, this thrust has not yet delivered any definitive (or, for that matter, unexpected) results pertaining to the documented departures from the conventional Fermi liquid (FL). 

Another school of thought followed the 'tried-and-true' conventional diagrammatics, mostly seeking to justify - even if merely quantitatively (in the absence of any suitable parameter) -  
the customary Migdal-Eliashberg approximation that neglects vertex corrections \cite{chub}. 
By its very nature this approach would seem to be limited to the regime where the interactions - while potentially important - would not yet become dominant.

Alternatively, a number of authors 
has entertained the 'let's just talk about it' type of approach, 
thus bringing into existence such constructs as 'post-modern' and 'ersatz' FL \cite{talk}. Many of those essays tend to feature only a few equations that appear to be either some symmetry transformations or the (supposedly, equivalent) re-writings of the original Hamiltonian, much of the discussion revolving around such trendy concepts as topology, anomalies, fractons, etc. 

Taken at their face value, such analyses tend to conform to the various hydrodynamic 
descriptions, also encroaching into the territory claimed by the late-stage applied 
(a.k.a. 'bottom-up' or 'AdS/CMT') holography \cite{ads}.
In turn, the latter, after having remained the (purportedly) 'well established ultimate solution to every strongly correlated problem' for the past $15+$ years has been 
finally sinking under the radar, as of lately 
(akin to the fate of any other 'cargo cult science' \cite{cult}, the eventual demise of AdS/CMT should have been anticipated \cite{mystery}), except for still using its recognizable 
brand name in the 'safer-haven' field of (re)formulating and extending the conventional hydrodynamics.  

Complimentary to the above approaches, similar goals have also been pursued in the framework of the so-called multi-dimensional bosonization that aims at substituting the underlying fermionic description with some effective bosonic one. As a common element to all the early constructions, 
a Fermi surface (FS) of the $d>1$-dimensional fermions would be divided onto a collection of small 'patches',
so that the fermions with momenta belonging to one patch would then be 
viewed as (pseudo) one-dimensional and treated by virtue of the conventional $1d$ bosonization \cite{bos}.   
However, despite the initial enthusiasm the status of the early explorations has long remained unsettled.   
One of the subtle - but often ignored - issues has been a potentially
 important role of finite FS curvature \cite{curvature}. 

Parallel to - and inspired by - the above efforts, there were also some (apparently, little-noticed) attempts to formulate a fully geometric bosonization procedure based on the Kirillov-Kostant (KK) 
method of coadjoint orbit quantization
and formulated as a path integral over the Wigner function-type field variable \cite{geom}. 

Incidentally, after its more recent exposition in the context of an alternate holography-like analysis of the phase-space dynamics \cite{hydro}, the KK method was resurrected (with only casual  
and out-of-context mentioning of Refs.\cite{geom,hydro} - or avoiding it altogether) 
in a growing number of publications \cite{kk}. Albeit belatedly, 
the subject has become popularized and finally caught in the public eye.
In particular, it was meticulously demonstrated to reproduce, in its linearized form, the standard FL description. However, the full potential of this promising technique still remains to be ascertained
and its applications are yet to be extended beyond the FL realm,
as its only practical use in such capacity so far dates back to the little-known Refs.\cite{geom,hydro}.

Regardless of all the differences between the various takes on multi-dimensional bosonization, 
at the core of such constructions is the quadratic action   
\bea
S=\int_{t,{\bf r}}\oint_{\bf n}
({\bf n}{\bf \nabla})\phi_{\bf n}
({\partial\over \partial t}\phi_{\bf n}-v({\bf n}{\bf \nabla})\phi_{\bf n})\nonumber\\
-{v\over 2}\int_{t,{|bf r}}\oint_{\bf n}\oint_{\bf n^{\prime}}
({\bf n}{\bf \nabla}\phi_{\bf n})
{F}({\bf n},{\bf n^{\prime}})
({\bf n^{\prime}}{\bf \nabla}\phi_{\bf n^{\prime}})
\eea
in terms of the 'patch' bosonic field $\phi_{\bf n}$ which represents fluctuations of fermion density,  
$\delta\rho(t,{\bf r})=\oint_{\bf n}({\bf n}{\bf \nabla})\phi_{\bf n}+\dots$, 
and is labeled by the unit normal vector $\bf n$ -
or, equivalently, $d-1$ angular variables (in 2d the only one, $\theta$) - parametrizing a continuous FS.
The Fermi velocity $v$ in (1) is characteristic of a generic ('non-flat') energy band. 

As compared to (1), in the KK approach the exact bosonized action includes, both, the 
interaction and Berry phase (a.k.a., WZW) parts, thus featuring additional
cubic, quartic, and still higher orders of the gradients of $\phi_{\bf n}(t,{\bf r})$ \cite{geom,hydro,kk}.   

To first order in the (off-diagonal) quadratic Landau kernel ${F}({\bf n},{\bf n^{\prime}})$ 
which accounts for the quasiparticle renormalization effects
the bosonic propagator reads 
\be
<\phi_{\bf n}\phi_{\bf n^{\prime}}>
={\delta({\bf n}-{\bf n}^{\prime})\over ({\bf n}{\bf q})(\omega-v({\bf n}{\bf q}))}+
{F({\bf n},{\bf n}^{\prime})\over 
(\omega-v({\bf n}{\bf q}))(\omega-
v({\bf n}^{\prime}{\bf q}))}
\ee
The task of computing this function beyond the linear order  
amounts to solving the kinetic equation
\be 
(\omega-v({\bf n}{\bf q}))\phi_{\bf n}
-v({\bf n}{\bf q})\oint_{n^{\prime}} 
F({\bf n},{\bf n}^{\prime}) 
({\bf n^{\prime}}{\bf q})\phi_{\bf n^{\prime}}=I[\phi_{\bf n}]
\ee
where the r.h.s. represents a collision integral. 

In the conventional FL, solutions to the collisionless kinetic equation
describe a continuum of particle-hole excitations for $\omega<vq$
and their bound states for $\omega>vq$, while 
the collision term provides the latter with a finite width $\gamma$. 

It has long been argued that Eq.(3) can be applied even 
in the absence of well-defined quasiparticles,
provided that the single-fermion Green function 
remains sharp as a function of the normal component of momentum $p_{\parallel}$ 
in the vicinity of a fiducial FS. 
Specifically, Eq.(3) can still be derived by using the Keldysh 'lesser' function 
integrated over the quasiparticle energy $\xi_p=v p_{\parallel}$  \cite{qbe}, 
$
f(t, {\bf r}; \epsilon, \xi_p,{\bf n})=\int {d\xi_p\over 2\pi i} G_{<}(\epsilon, {\bf p} |t, {\bf r})
$. By further integrating over frequency one obtains an angular-resolved 
variation of the chemical potential 
$
\rho_{\bf n}(t, {\bf r})=\int {d\epsilon\over 2\pi} 
f(t, {\bf r}; \epsilon, \xi_p,{\bf n})
$
which measures a local FS displacement and sums up to the total density fluctuation,
$\delta\rho(t,{\bf r})=\oint_{\bf n}\rho_{\bf n}(t, {\bf r})$.  

The present note addresses a previously overlooked complication that one  
encounters while using the bosonization approach even in the linearized regime. 
 The origin of this subtlety is rooted in the peculiar 2d kinematics of particle-hole excitations. 
Namely, in the case of a convex (albeit not a concave)
$2d$ FS one finds that even in the FL regime there exists a strong disparity
between the scattering rates $\gamma_{\pm}$ 
of the even and odd angular harmonics of the local density
$\rho_{\theta}(t, {\bf r})=\sum_l\rho_l({\bf r},t)e^{il\theta}$ 
expanded in the angular momentum basis \cite{tomo}.  

This behavior was predicted for any Landau function $F(\theta,\omega)$ 
with a non-singular dependence on the scattering angle 
$\theta=\cos^{-1}({{{\bf n}{\bf n^{\prime}}}})$, whose 
harmonics $F_l$ are determined by $\theta\sim 1$
and show no singularity at vanishing $\omega$. 
However, the situation becomes further involved in those cases  
where the Landau function is singular 
at small $\theta$ and, therefore, must retain some $\omega$-dependence 
in order to yield a finite $F_l$.   

One plausible origin of the NFL 
states has long been identified as the long-ranged  and/or retarded interactions. 
Many of, both, established and conjectured NFL fall into the universality class of 
non-relativistic fermions coupled via an overdamped bosonic (e.g., gauge) 
field with the dispersion relation $\omega\sim iq^z$ controlled by a dynamical exponent $z>1$
(hereafter, without any special tuning between the competing parameters 
all the momenta/frequencies are measured in units of the Fermi momentum $k_F$ and energy 
$vk_F$). 

In this scenario,  the quasiparticle interaction function 
\be
F(\theta, \omega) = {|\theta|\over |\theta|^z+\omega}
\ee
gives rise to the strongly energy-dependent fermion self-energy 
\be
\Sigma(\omega)\sim\omega^{d/z}
\ee
which behavior has been extensively studied in a variety of the purported 'strange metals',
including itinerant ferro-magnets, 
electromagnetic response in ordinary metals, spinon gauge theories of spin liquids,  
compressible QHE, Ising nematics and other Pomeranchuk/Lifshitz transitions,
or even the (non-)abelian quark-gluon plasma \cite{gauge}.  

In a system governed by the Landau function (4), scattering is predominantly 
small-angle and characterized by a typical energy transfer $\omega$  
being small compared to the momentum normal to 
the FS - which, in turn, is small compared to the tangential one at low energies
for all $z>d$
\be 
\omega\ll q_{\parallel}\sim\omega^{d/z} \ll q_{\perp}\sim\omega^{1/z}
\ee
Notably, in this regime scattering becomes 
primarily momentum, rather than energy, dependent, hence quasi-elastic.

A recent follow-up of the original work \cite{qbe} studied the 
possibility of undamped zero-sound and other 
collective modes in the NFL regime.
Different takes on the topic of the existence of such modes either supported, refuted, 
or found the answer to depend on the interaction strength  \cite{mandal}.
In particular, it was argued that an undamped zero-sound mode may still exist in 
the weak coupling limit, whereas at strong couplings it is likely to be buried 
inside the particle-hole continuum. 
 
However, the studies \cite{mandal} were either limited to the collisionless limit ($\gamma\to 0$)
or ignored the important cancellations occurring between the quasiparticle 
self-energy $\Sigma$ and the Landau function $F$, as required by energy/momentum conservation. 
Moreover, those analyses were oblivious to the
aforementioned disparity between the relaxation rates of 
the even and odd harmonics in the case of a convex (albeit not a concave)
FS. First reported in the conventional FL \cite{tomo} 
this prediction was also extended onto the recent explorations of 
the 'Ising-nematic' variant of the problem (4) \cite{guo} and beyond \cite{strange}.

The crux of the matter is that while the even-$l$ harmonics can be relaxed solely by 
the shape fluctuations of an incompressible FS (hence, the corresponding amplitudes acquire 
extra powers of $q_{\perp}$), the odd-$l$ ones 
require changes to the FS volume (hence, the corresponding 
amplitudes involve powers of $q_{\parallel}$). 
This kinematic property gives rise to a delayed onset of angular equilibration, as opposed to the 
uni-directional relaxation, thus resulting in the emergence 
of a new, 'tomographic', regime intertwining between
the ballistic (collisionless) and ordinary hydrodynamic (diffusive) ones \cite{tomo}.  

Based on the $2d$ parity ($\theta\to\pi+\theta$) and particle-hole ($\xi\to -\xi$) symmetries
the pertinent scattering rates must be proportional to the even powers of $q_{\perp,\parallel}$
(i.e., of the scattering angle $\theta\sim\omega^{1/z}$) as well as the matching powers of the conjugate momentum $l$, thus resulting in the phenomenological dependencies \cite{strange}
\be
\gamma_+(\omega,l)\sim l^{2n}\omega^{d+2n\over z},~~~
\gamma_-(\omega,l) \sim l^{2n+4m}\omega^{d+2n+2dm\over z}
\ee
for the even and odd angular harmonics, respectively. 
Although (7) pertains to the rotationally-invariant FS,
conceivably, the additional suppression of the odd rates is expected to still hold 
for a generic convex FS.

At the momenta in excess of 
$
l_{\perp}=\omega^{-1/z}
$
the small-angle suppression in Eqs.(7) ceases to exist. 
Furthermore, at the momenta higher than  
$
l_{\parallel}=\omega^{-d/2z}
$
the faster-growing odd rate catches up with the even one.  
Notably, in $d=2$ the two scales merge into one and same, $l_{\perp}=l_{\parallel}=l_0$,
above which both rates approach their common asymptotic value 
$\gamma_{\pm}(l\gtrsim l_0)\sim\Sigma(\omega)$.  

To keep the discussion as general as possible
one can also allow for additional suppression at low $\omega$ through the matrix elements in (7)   
for $n,m\geq 0$. Alongside the index $z$, the integers $n$ and $m$ characterize 
the 'universality class' of the NFL in question.   

The combined effects of the Landau function (spectrum renormalization) 
and collision term (excitation width) can be accounted for
by introducing an overall (complex-valued) bosonic self-energy 
$\sigma_{\pm}(\omega,l)=\omega(F_0-F_{l})+i\gamma_{\pm}$.
In the (un)conventional $2d$ FL with predominantly 
large-angle scattering and $Re\Sigma(\omega)\sim\omega\gg Im\Sigma(\omega)\sim\omega^2$ 
one has to choose $z=1$, $n=0$, and $m=1$, thus obtaining   
\be
\sigma_{+}(\omega,l)\sim \omega+ i\omega^2,~~~
\sigma_{-}(\omega,l)\sim \omega^3 + il^4\omega^4
\ee
Upon a closer inspection, the imaginary term in $\sigma_{-}$ acquires an additional 
logarithmic factor $\ln l$ originating from the kinematic divergence specific to 2d \cite{tomo}.

By contrast, in any genuine NFL with a sub-linear self-energy ($z>1$) 
the real and imaginary parts of $\sigma_{\pm}$ appear to be of comparable magnitude. 
In particular, in the extensively studied 
problem of the overdamped 2d bosonic mode with $z=3$ and $n=m=1$  
one finds \cite{strange,guo}
\be 
\sigma_{+}(\omega,l)\sim l^2\omega^{4/3},~~~~~~
\sigma_{-}(\omega,l)\sim l^6\omega^{8/3}
\ee 
With the input of Eq.(7), assuming a nearly isotropic fermion dispersion 
and expanding over the angular harmonics 
one converts Eq.(3) into a pair of coupled equations 
\bea
(\omega+\sigma_+(\omega,l))\rho_+ - vq\cos\theta\rho_- = 0\nonumber\\
(\omega+\sigma_-(\omega,l))\rho_- - vq\cos\theta\rho_+ = 0
\eea  
The density harmonic 
$\rho_l$ oscillates strongly between the even and odd angular momenta, thereby disallowing one from 
treating it as a smooth function of $l$ and performing a naive gradient expansion.
Instead, similar to the effective low-energy theory of antiferromagnets, 
a  long-wavelength description should involve two 
separate fields  $\rho_{\pm}$ composed solely of the even and odd harmonics, respectively.
Moreover, the even component appears to be 
still faster then its odd counterpart, as per the relaxation 
rates (7).  

Eliminating the 'fast' variables 
$\rho_+$ in favor of the 'slow' $\rho_-$ yields a closed equation for the latter 
\be
[(\omega+\sigma_{+}(\omega,l))(\omega+\sigma_{-}(\omega,l))-
(vq\cos\theta)^2]\rho_-=0
\ee
In the continuum limit, by further expanding (11) over $\theta$ and quantizing 
the resulting expression, $\theta\to -i\partial_l$,  one arrives at the effective $2nd$ order 
differential eigen-state equation 
\be
[{V(l)\over (vq)^2}-\partial_l^2)]\rho_-=\lambda\rho_-
\ee
with the eigen-value $\lambda=(\omega/vq)^2-1$ and the effective $1d$ potential 
\be
V(l)=(\sigma_+(\omega,l)+\sigma_-(\omega,l))\omega+\sigma_+(\omega,l)\sigma_-(\omega,l)
\ee
which alternates with $l$ between the two power-law asymptotics appearing in Eq.(7). 
It is worth noting that, had it not been for the 'spinor' (two-component) nature of Eqs.(10) 
the $1d$ potential (13) would have been given by a simple sum of $\sigma_{\pm}$. 

In the semiclassical approximation Eq.(12) produces a quantization condition  
\be
\int^{l_m}_{-l_m} dl(\lambda_N-{V(l)\over (vq)^2})^{1/2}=\pi (N+\delta)
\ee
where $\delta\sim 1$ and 
$l_m$ is the turning point given by the equation $V(l_m)=\lambda_N(vq)^2$.

At $l\gtrsim l_0$ the potential (13) levels off, thus suggesting that 
the discrete eigen-states (if any) 
should be confined to the range of momenta $l< l_0$, so that their overall 
number would be bounded from above by $N_{max}\approx l_0$. 
 
For small momenta ($l<<l_0$), the effective potential is dominated by the first term in (13) 
$
V_<(l)\sim l^{2n}\omega^{(d+z+2n)/z}$
which attains its maximal value $V_<(l\gtrsim l_0)\sim\omega^{(d+z)/z}$
at large $l$.
In contrast, while initially subdominant, the last term in (13), 
$
V_>(l)\sim l^{4(n+m)}\omega^{(2d+4n+4m)/z}
$, grows faster and for $l\gtrsim l_0$
flattens off at the value $V_>(l\gtrsim l_0)\sim\omega^{2d/z}$. 
The latter appears 
to be higher than $V_<(l\gtrsim l_0)$ for $z>d$.
A crossover between the two competing contributions occurs at 
$l_1=\omega^{(z-d-2n-4m)/z(2n+4m)} << l_0$. 

Solving for the eigen-values $\lambda_N$ one finds 
the spectra of 'zero-sound' collective modes. At low $\omega$ those read    
\be
\omega_N = \pm vq + C^<_Nq^{\nu_<}
\ee 
where
\be
C^<_N\sim N^{2n\over n+1},~~~\nu_<={d+n(z+2)\over z(n+1)}
\ee 
The two solutions in (15) correspond to the modes of opposite chirality, akin to 
the spectrum of antiferromagnons defined in the halved Brillouin zone.

At low $\omega$ and $N\ll N_{max}$  
the $2nd$ (non-linear) term in the r.h.s. of (15) remains small compared to the linear one, 
thus resulting in the sub-dominant 
upward deviation from the upper boundary of the particle-hole continuum.

For the aforementioned parameters $n=m=1$, $d=2$, and $z=3$
one finds that non-linear term in (15) scales as $N\omega^{7/6}$. 
However, as $N$ increases towards $N_{max}$ 
the non-linear term in (15) grows to values of order $\omega^{5/6}$, thus suggesting that the  
dispersion relations of the modes with $N\sim N_{max}$ tend to become  
essentially non-linear and/or strongly damped.

However, as a more accurate description,  
for $l_1\lesssim l\lesssim l_0$ the potential (13) would be governed by $V_>(l)$, thus  producing 
a different set of the eigen-values $\lambda_N$, so that the counterpart of Eq.(16) should then read 
\be
C^>_N\sim N^{(4n+4m)\over 2n+2m+1},~~
\nu_> = {2d-z+2(z+2)(n+m)\over z(1+2n+2m)}
\ee
For the parameter values corresponding to the Ising nematic scenario
the asymptotic low-$q$ behavior of the non-linear term in
Eq.(15) is  $N^{8/5}\omega^{6/5}$. At the largest values of $N\sim N_{max}$ it 
approaches $\omega^{2/3}$, in agreement with the behavior of the full-fledged fermion self-energy (5).  

It should also be mentioned that the above power-law asymptotics are not
directly applicable to the standard FL, 
as the real and imaginary parts of $\Sigma(\omega)$ scale differently in this case. However, by adjusting the above analysis and choosing $n=0$, $m=1$, $d=2$, and $z=1$, one finds that at low energies the resulting
collective modes' dispersion relations 
\be
\omega_N = \pm (1+A_N)vq +i B_Nq^{2}
\ee
with $A_N\sim B_N\sim N^{4/3}$ remain linear and weakly damped.
The actual number of such modes  is going to be limited 
by the threshold criteria which guarantee a positivity of the prefactor in the linear term in (18))
and which restrict the (model-specific) values of the Landau parameters $F_l$. 

In the NFL case, the singular nature of $F_l(\omega)$ and the concomitant
increase in the number of the collective modes ($N_{max}\to\infty$) at $\omega\to 0$  
suggest that an increasingly large 
number of such modes may accumulate near the upper edge  
of the naive particle-hole continuum. 
This picture is consistent with the interpretation of the collective modes with low $N$ 
as smooth FS oscillations  that are delocalized over the entire FS. 
The strong quasiparticle renormalization effects signified by the fermion self-energy (5) do not manifest themselves in the spectra of such weakly-damped modes. 

On the contrary, rough local FS fluctuations representing the unbound particle-hole pairs 
involve large $l$, thereby exhibiting a (potentially singular) fermion self-energy.
Their spectral relations are non-linear and damped, 
likely corresponding to the non-hydrodynamic modes that would be unobtainable by virtue 
of a naive gradient expansion.

Taken at their face value, the above results suggest that in order to properly incorporate the above 
collective mode spectra, a consistent effective low-energy 
description would need to start out with the quadratic bosonic action
\be
S_{eff}={1\over 2}\sum_N\int_{t,q}
(({\partial\over \partial t}\phi_N)^2-
C^2_N(vq)^{2(1+\nu)}\phi_N^2)
\ee
which is reminiscent of the theory of mobile (non-chiral) fracton excitations. 
It is worth reiterating that the quadratic form (19) gets naturally diagonalized in the space of 
quantum numbers of the effective $1d$ potential (13), rather than the basis of angular momenta. 

The exotic properties of the NFL collective modes, including pertinent energy, momentum, and temperature dependencies, can be probed in the non-local (finite momentum) transport, 
bi-layer drag, pump-probe, and other experiments. For one, a non-local conductivity was predicted to 
exhibit certain 'strange metal'-like features - including such an (ostensibly) unifying   
hallmark of the NFL behavior as linear resistivity - even in the ordinary FL \cite{tomo}.
    
On the theory side, making concrete experimental predictions requires a systematic calculation
of the fully dressed  charge/current response functions, complete with their poles and branch cuts, both, at small momenta and close to $2k_F$. 
In that regard, to access the non-trivial regimes beyond the RPA routine one might need to combine the 
KK approach to multi-dimensional bosonization with 
a further advanced eikonal-like non-perturbative technique of Refs.\cite{hydro,stamp}.

This work was performed in part at Aspen Center for Physics, 
which is supported by National Science Foundation grant PHY-2210452.


\end{document}